\documentclass[useAMS,usenatbib]{mn2e}
\usepackage{epsfig}
\usepackage{amsmath}

\newcommand{\be}{\begin{equation}}
\newcommand{\beq}{\begin{equation}}
\newcommand{\ba}{\begin{eqnarray}}
\newcommand{\ee}{\end{equation}}
\newcommand{\eeq}{\end{equation}}
\newcommand{\ea}{\end{eqnarray}}

\newcommand{\hs}{\hspace{1mm}}

\newcommand{\apj}{ApJ}
\newcommand{\aap}{A\&A}
\newcommand{\apjl}{ApJL}
\newcommand{\mnras}{MNRAS}
\newcommand{\aj}{AJ}
\newcommand{\apjs}{ApJS}
\newcommand{\nat}{{\it Nature}}
\newcommand{\araa}{ARA\&A}
\newcommand{\pasj}{PASJ}

\def\lsim{~\rlap{$<$}{\lower 1.0ex\hbox{$\sim$}}}

\def\gsim{~\rlap{$>$}{\lower 1.0ex\hbox{$\sim$}}}

\title[Very Massive Stars in High-z Galaxies]{Very Massive Stars in High-Redshift Galaxies}

\author[Mark Dijkstra \& J. Stuart B. Wyithe]{Mark Dijkstra$^{1}$\thanks{E-mail:dijkstra@physics.unimelb.edu.au} and J. Stuart B. Wyithe$^{1}$\thanks{E-mail:swyithe@physics.unimelb.edu.au}\\
$^{1}$School of Physics, University of Melbourne, Parkville, Victoria, 3010, Australia}
\def\LaTeX{L\kern-.36em\raise.3ex\hbox{a}\kern-.15em
    T\kern-.1667em\lower.7ex\hbox{E}\kern-.125emX}

\begin{document}

\date{\today}
\pagerange{\pageref{firstpage}--\pageref{lastpage}} \pubyear{2006}

\maketitle

\label{firstpage}
\begin{abstract}
A significant fraction of Ly$\alpha$ emitting galaxies (LAEs) at $z\geq 5.7$ have rest-frame equivalent widths (EW) greater than $\sim100$\AA. However only a small fraction of the Ly$\alpha$ flux produced by a galaxy is transmitted through the IGM, which implies intrinsic Ly$\alpha$ EWs that are in excess of the maximum allowed for a population-II stellar population having a Salpeter mass function. In this paper we study characteristics of the sources powering Ly$\alpha$ emission in high redshift galaxies. We propose a simple model for Ly$\alpha$ emitters in which galaxies undergo a burst of very massive star formation that results in a large intrinsic EW, followed by a phase of population-II star formation with a lower EW. We confront this model with a range of high redshift observations and find that the model is able to simultaneously describe the following eight properties of the high redshift galaxy population with plausible values for parameters like the efficiency and duration of star formation: i-iv) the UV and Ly$\alpha$ luminosity functions of LAEs at z=5.7 and 6.5,  v-vi) the mean and variance of the EW distribution of Ly$\alpha$ selected galaxies at z=5.7, vii) the EW distribution of i-drop galaxies at z$\sim$6, and viii) the observed correlation of stellar age with EW. Our modeling suggests that the observed anomalously large intrinsic equivalent widths require a burst of very massive star formation lasting no more than a few to ten percent of the galaxies star forming lifetime. This very massive star formation may indicate the presence of population-III star formation in a few per cent of i-drop galaxies, and in about half of the Ly$\alpha$ selected galaxies.
\end{abstract}

 
\section{Introduction}
\label{sec:intro}
Narrow band searches for redshifted Ly$\alpha$ lines have discovered a large number of Ly$\alpha$ emitting galaxies with redshifts between $z=4.5$ and $z=7.0$ \citep[e.g.][]{Hu96,Hu02,MR02,Ko03,Dawson04,Hu04,Stanway04,Taniguchi05,Westra06,Ka06,Shima06,Iye06,Stanway07,Tapken07}. The Ly$\alpha$ line emitted by these galaxies is very prominent, often being the only observed feature. The prominence of the Ly$\alpha$ line is quantified by its equivalent width (EW), defined as the total flux of the Ly$\alpha$ line, $F_{{\rm Ly}\alpha}$ divided by the flux density of the continuum at 1216 \AA: EW$\equiv F_{{\rm Ly}\alpha}/f_{1216}$. Throughout this paper we refer to the rest-frame EW of the Ly$\alpha$ line (which a factor of $(1+z)$ lower than the EW in the observers frame). 

Approximately $50\%$ of Ly$\alpha$ emitters (hereafter LAEs) at $z=4.5$ and $z=5.7$ have lines with EW$\sim 100-500$ \AA\hs \citep{Dawson04,Hu04,Shima06}. For comparison, theoretical studies conclude that the maximum EW which can be produced by a conventional population of stars is 200-300 \AA. Moreover, this maximum EW can only be produced during the first few million years of a starburst, while at later times the luminous phase of Ly$\alpha$ EW gradually fades \citep{CF93,MR02}. Therefore, observed EWs lie near the upper envelope of values allowed by a normal stellar population.

The quoted value for the upper envelope of EW$\sim200-300$ \AA\hs corresponds to the emitted Ly$\alpha$ flux. However not all Ly$\alpha$ photons are transmitted through the IGM, and we expect some attenuation. Within the framework of a Cold Dark Matter cosmology, gas surrounding galaxies is significantly overdense, and possesses an infall velocity relative to the mean IGM \citep{Barkana-Infall}. As a net result, the IGM surrounding high redshift galaxies is significantly opaque to Ly$\alpha$ photons. Indeed it can be shown that for reasonable model assumptions, only $\sim 10-30\%$ of all Ly$\alpha$ photons are transmitted through the IGM \citep{IGM}. As a result, the intrinsic Ly$\alpha$ EW emitted by high redshift LAEs is systematically larger than observed. Indeed, this observation suggests that a significant fraction of LAEs at $z\geq4.5$ have intrinsic EWs that are much larger than can possibly be produced by a conventional population of young stars.

One possible origin for this large EW population is provided by active galactic nuclei (AGN), which can have much larger EWs due to their harder spectra \citep[e.g.][]{CF93}. However, large EW LAEs are not AGN for several reasons: (1) the Ly$\alpha$ lines are too narrow \citep{Dawson04} (2) these objects typically lack high--ionisation state UV emission lines, which are symptomatic of AGN activity \citep{Dawson04}, and (3) deep X-Ray observations of 101 Ly$\alpha$ emitters by Wang et al. (2004, also see Malhotra et al. 2003, Lai et al, 2007) revealed no X-ray emission neither from any individual source, nor from their stacked X-Ray images.

Several recent papers have investigated the stellar content of high-redshift LAEs by comparing stellar synthesis models with the observed broad band colors \citep{Finkelstein07}. These comparisons are often aided by deep IRAC observations on {\it Spitzer} \citep{Lai07,Pirzkal07}. In this paper we take a different approach. Instead of focusing on individual galaxies, our goal is to provide a simple model that describes the population of Ly$\alpha$ emitting galaxies as a whole. This population is described by the rest-frame ultraviolet (UV) and Ly$\alpha$ luminosity functions (LFs) at $z=5.7$ and $z=6.5$ , and the Ly$\alpha$ EW distribution at $z=5.7$ \citep{Shima06,Ka06}. The sample of high-redshift LAEs is becoming large enough that meaningful constraints can now be placed on simple models of galaxy formation.

The outline of this paper is as follows:
In \S~\ref{sec:model} -\S~\ref{sec:popIII} we describe our models. In \S~\ref{sec:discuss} we discuss our results, and compare with results from stellar synthesis models, before presenting our conclusions in \S~\ref{sec:conclusion}. The parameters for the background cosmology used throughout this paper are $\Omega_m=0.24$, $\Omega_{\Lambda}=0.76$, $\Omega_b=0.044$, $h=0.73$ and $\sigma_8=0.74$ \citep{Spergel06}.

\section{The Model}
\label{sec:model}
\citet{LF} found that the observed Ly$\alpha$ LFs at $z=5.7$ and $z=6.5$ are well described by a model in which the Ly$\alpha$ luminosity of a galaxy increases in proportion to the mass of its host dark matter $M_{\rm tot}$. One can constrain quantities related to the star formation efficiency from such a model \citep[also see][]{Mao07,Stark07}. 

However, it is also possible to obtain constraints from the rest-frame UV-LFs. In contrast to the Ly$\alpha$ LF, the UV-LF is not affected by attenuation by the IGM, which allows for more reliable constraints on quantities related to the star formation efficiency. In the first part of this paper (\S~\ref{sec:modelLF}-\S~\ref{sec:igm}) we present limited modeling to illustrate parameter dependences, using the UV-LF to constrain model parameters related to star formation efficiency and lifetime. These model parameters may then be kept fixed, and the Ly$\alpha$ LFs and EW distributions used to constrain properties of high redshift LAEs such as their intrinsic Ly$\alpha$ EW and the fraction of Ly$\alpha$ that is transmitted through the IGM. Later, in \S~\ref{sec:popIII}, we present our most general model, and fit to both the UV and Ly$\alpha$ LFs, as well as the EW distribution, simultaneously, treating all model parameters as free.

\section{Modeling the UV and Ly$\alpha$ Luminosity Functions.}
\label{sec:modelLF}
\subsection{Constraints from the UV-LF.}
\label{sec:uv}
\begin{figure}
\vbox{\centerline{\epsfig{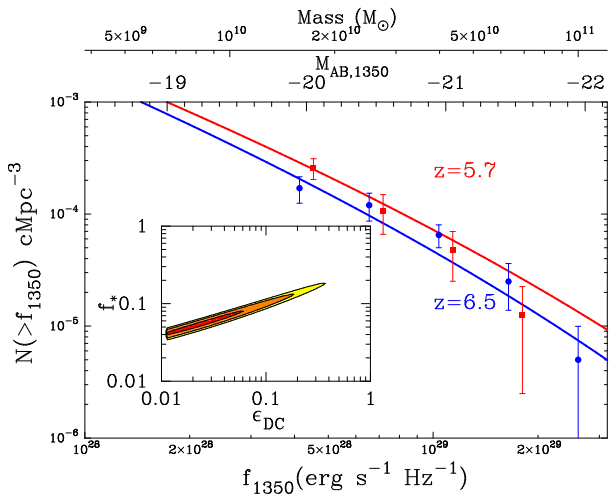}}}
\caption[]{Constraints on the star formation efficiency from the observed rest-frame UV-luminosity functions of LAEs at $z=5.7$ ({\it red squares}) and $z=6.5$ ({\it blue circles}) \citep{Shima06,Ka06}. In our best-fit model, a faction $f_*\sim 0.06$ of all baryons is converted into stars over a time-scale of $\epsilon_{\rm DC}  t_{\rm hub}\sim 0.03$ Gyr (see text). The inset shows likelihood contours in the ($\epsilon_{\rm DC},f_*$)-plane at $64\%$, $26\%$ and $10\%$ of the peak likelihood. Also shown on the upper horizontal axis is the mass corresponding to $M_{\rm AB,1350}$ in the best-fit model.}
\label{fig:uv}
\end{figure}

We begin by presenting a simple model for the UV-LF \citep{WL07,Stark07}. In Figure~\ref{fig:uv} we show the rest-frame UV-LFs of LAEs at $z=5.7$ and $z=6.5$ \citep{Shima06,Ka06}. We use the following simple prescription to relate the ultraviolet flux density emitted by a galaxy, $f_{1350}$, to the mass of its host dark matter $M_{\rm tot}$. The total mass of baryons within a galaxy is $(\Omega_b/\Omega_m)M_{\rm tot}$, of which a fraction $f_*$ is assumed to be converted into stars over a time scale of $t_{\rm sys}=\epsilon_{\rm DC}t_{\rm hub}$. Here, $\epsilon_{\rm DC}$ is the duty cycle and $t_{\rm hub}(z)$, the Hubble time at redshift $z$. This prescription yields a star formation rate of $\dot{M}_*=f_*(\Omega_b/\Omega_m)M/t_{\rm sys}$. The star formation rate can then be converted into $f_{1350}$ using the relation $f_{1350}=7 \times 10^{27}(\dot{M}_*/[M_{\odot}/{\rm yr}])$ erg s$^{-1}$ Hz$^{-1}$ \citep{K98}. The precise relation is uncertain but differs by a factor of less than 2 between a normal and a metal-free stellar population (see e.g. Loeb et al. 2005). Uncertainty in this conversion factor does not affect our main conclusions. The presence of dust would lower the ratio of $f_{1350}$ and $\dot{M}_*$, which could be compensated for by increasing $f_*$. However, \citet{Bouwens06} found that dust in $z=6.0$ Lyman Break Galaxies (LBGs) attenuates the UV-flux by an average factor of only $1.4$ (and dust obscuration may be even less important in LAEs, see \S~\ref{sec:q}). Since this is within the uncertainty of the constraint we obtain on $f_*$, we ignore extinction by dust. The number density of LAEs with UV-flux densities exceeding $f_{\rm 1350}$ is then given by
\begin{equation}
N(>f_{\rm 1350})=\epsilon_{\rm DC}\int_{M_{\rm UV}}^{\infty}dM\frac{dn}{dM},
\label{eq:form} 
\end{equation} where $M_{\rm UV}$ is the mass that corresponds to the flux density, $f_{\rm 1350}$ (through the relations given above). The function $dn/dM$ is the Press-Schechter (1974) mass function \citep[with the modification of][]{ST}, which gives the number density of halos of mass $M$ (in units of comoving Mpc$^{-3}$)\footnote{Our model effectively states that the star formation rate in a galaxy increases linearly with halo mass. This is probably not correct. To account for a different mass dependence we could write the star formation rate as $\dot{M}_*\propto M^{\beta}$, where $\beta$ is left as a free parameter. However, the range of observed luminosities span only 1 order of magnitude, and we will show that the choice $\beta=1$ provides a model that describes the observations well. Furthermore, the duty cycle $\epsilon_{\rm DC}$ may be viewed as the fraction of dark matter halos that are currently forming stars. The remaining fraction ($1-\epsilon_{\rm DC}$) of halos either have not formed stars yet, or are evolving passively. In either case, the contribution of these halos to the UV-LF is set to be negligible.}. The free parameters in our model are the duty cycle, $\epsilon_{\rm DC}$, of the galaxy, and the fraction of baryons that are converted into stars, $f_*$. We calculated the UV-LF for a grid of models in the ($\epsilon_{\rm DC},f_*$)-plane, and generated likelihoods ${\mathcal L}[P]={\rm exp}[-0.5\chi^2]$, where $\chi^2=\sum_i^{N_{\rm data}}({\rm model}_i-{\rm data}_i)^2/\sigma^2_{i}$, in which data$_i$ and $\sigma_i$ are the i$^{\rm th}$ UV-LF data point and its error, and model$_i$ is the model evaluated at the $i^{\rm th}$ luminosity bin. The sum is over $N_{\rm data}=8$ data points. The inset in Figure~\ref{fig:uv} shows the resulting likelihood contours in the ($\epsilon_{\rm DC},f_*$)-plane at $64\%$, $26\%$ and $10\%$ of the peak likelihood. The best fit model has $(\epsilon_{\rm DC},f_*)=(0.03,0.06)$ and is plotted as the solid line. In the following sections we assume this combination of $f_*$ and $\epsilon_{\rm DC}$.

\subsection{Constraints from the Ly$\alpha$ LF.}
\label{sec:lya1}
We next model the Ly$\alpha$ LF, beginning with the best-fit model of the previous section. The number density of LAEs at redshift $z$ with Ly$\alpha$ luminosities exceeding $\mathcal{T}_{\alpha}\times L_{\alpha}$ is given by \citep{LF}
\begin{equation}
N(>\mathcal{T}_{\alpha}\times L_{\alpha},z)=\epsilon_{\rm DC}\int_{M_{\alpha}}^{\infty}dM\frac{dn}{dM}(z),
\label{eq:form} 
\end{equation} where the Ly$\alpha$ luminosity and host halo mass, $M_{\alpha}$ are related by
\begin{equation}
\mathcal{T}_{\alpha} \times L_{\alpha}=\mathcal{L}_{\alpha}\frac{M_{\alpha}(M_{\odot}) \frac{\Omega_b}{\Omega_m}f_*}{t_{\rm sys}({\rm yr})}\mathcal{T}_{\alpha}.
\label{eq:2}
 \end{equation} In this relation, $\mathcal{T}_{\alpha}$ is the IGM transmission multiplied by the escape fraction of Ly$\alpha$ photons from the galaxy, and $\mathcal{L}_{\alpha}=2.0 \times 10^{42}\hs{\rm erg} \hs{\rm s}^{-1}$/($M_{\odot}$ yr$^{-1}$), is the Ly$\alpha$ luminosity emitted per unit of star formation rate (in $M_{\odot}$ yr$^{-1}$). Throughout, $\mathcal{L}_{\alpha,42}$ denotes $\mathcal{L}_{\alpha}$ in units of $10^{42}\hs{\rm erg} \hs{\rm s}^{-1}/(M_{\odot}$ yr$^{-1}$). We have taken $\mathcal{L}_{\alpha,42}=2.0$, which is appropriate for a metallicity of $Z=0.05Z_{\odot}$ and a Salpeter IMF (Dijkstra et al, 2007). Note that when comparing to observed luminosities \citep{Shima06,Ka06}, we have replaced $L_{\alpha}$ with $\mathcal{T}_{\alpha}\times L_{\alpha}$. This is because the observed luminosities have been derived from the observed fluxes by assuming that all Ly$\alpha$ emerging from the galaxy was transmitted by the IGM, whereas there is substantial absorption \citep[e.g.][]{IGM}. The product $\mathcal{T}_{\alpha}\times L_{\alpha}$ may be written as $\mathcal{T}_{\alpha}\times L_{\alpha}=4\pi d^2_L(z)S_{\alpha}$, where $S_{\alpha}$ is the total Ly$\alpha$ flux detected on earth and $d_L(z)$ is the luminosity distance to redshift $z$. The product $\mathcal{T}_{\alpha}\times L_{\alpha}$ may therefore be viewed as an effective luminosity inferred at earth. Furthermore, the selection criteria used by \citet{Shima06} and \citet{Ka06} limits these surveys to be sensitive to LAEs with EW$\gsim 10$\AA. In the reminder of this paper, the EW of model LAEs is always larger than this EW$_{\rm min}$, and we need not worry about selection effects when comparing our model to the data.

\begin{figure}
\vbox{\centerline{\epsfig{file=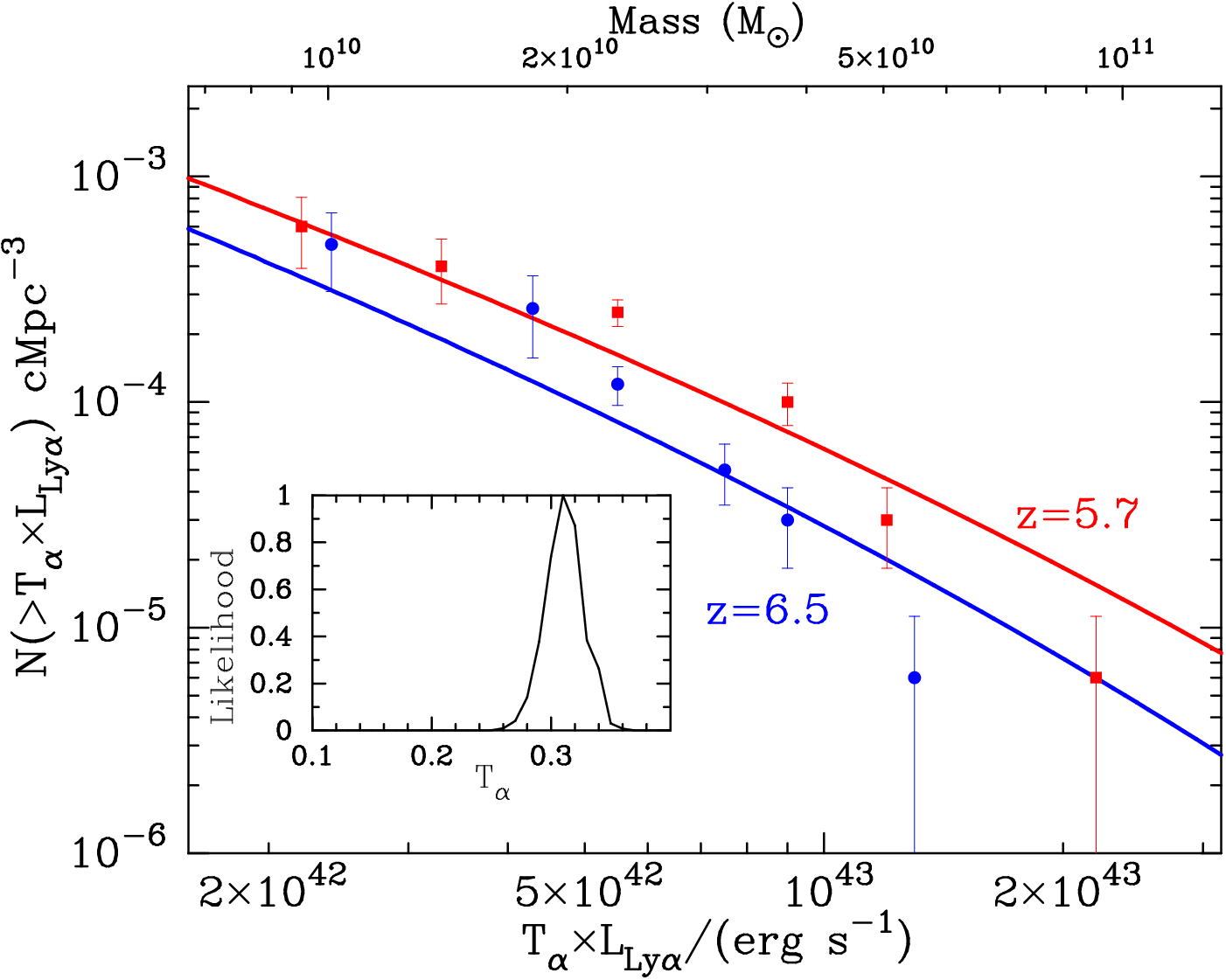,angle=270,width=8.0truecm}}}
\caption[]{Joint constraints on the IGM transmission from the observed Ly$\alpha$ and rest-frame UV LFs of LAEs at $z=5.7$ \citep{Shima06} and $z=6.5$ \citep{Ka06}. The {\it red squares} and {\it blue circles} represent the data at $z=5.7$ and $z=6.5$. Using the model that best describes the UV-LFs with $(f_{\rm star},\epsilon_{\rm DC})=(0.06,0.03)$ we fit to the observed Ly$\alpha$ LF. The only free parameter of our model was the fraction of Ly$\alpha$ that was transmitted through the IGM at $z=5.7$, $\mathcal{T}_{\alpha,57}$ (see text). Shown in the inset is the likelihood for $\mathcal{T}_{\alpha,57}$, normalised to a peak of unity. The figure shows that in order to simultaneously fit the Ly$\alpha$ and UV-LFs, only $\sim 30\%$ of Ly$\alpha$ photons are transmitted through the IGM. The best fit models are overplotted as the {\it solid lines}. }
\label{fig:lya1D}
\end{figure}

 In this section, we set the transmission at $z=6.5$ (denoted by $\mathcal{T}_{\alpha,65}$) to be a factor of $\sim 1.2$ lower\footnote{For our primary results in \S~\ref{sec:popIII} we allow this ratio to be a free parameter. The results presented in this section is not sensitive to the precise choice of the ratio of IGM transmission at $z=5.7$ and $z=6.5$.} than at $z=5.7$ (denoted by $\mathcal{T}_{\alpha,57}$). This ratio is the median of the range found by \citet{IGM}. We then calculated the Ly$\alpha$ LF for a range of $\mathcal{T}_{\alpha,57}$, and generated likelihoods $\mathcal{L}[P]={\rm exp}[-0.5\chi^2]$, where $\chi^2=\sum_i^{N_{\rm data}}({\rm model}_i-{\rm data}_i)^2/\sigma^2_{i}$, for each model. Here, data$_i$ and $\sigma_i$ are the i$^{\rm th}$ data point and its error, and model$_i$ is the model evaluated at the $i^{\rm th}$ luminosity bin. The sum is over $N_{\rm data}=6$ points at each redshift. In Figure~\ref{fig:lya1D} we show the Ly$\alpha$ luminosity functions at $z=5.7$ and $z=6.5$. The {\it red squares} and {\it blue circles} represent data from \citet[][$z=5.7$]{Shima06} and \citet[][$z=6.5$]{Ka06}, respectively.

The likelihood for $\mathcal{T}_{\alpha,57}$ (normalised to a peak of unity) is shown in the inset. The best fit model is overplotted as the {\it solid lines}, for which the value of the transmission is $\mathcal{T}_{\alpha,57}=0.30$. The modeling presented in this and the previous section therefore suggests that, in order to simultaneously fit the Ly$\alpha$ and UV luminosity functions, only $\sim 30\%$ of the Ly$\alpha$ can be transmitted through the IGM.  This transmission is in good agreement with the results obtained by \citet{IGM}, who modeled the transmission directly and found that for reasonable model parameters the transmission must lie in the range $0.1 \lsim \mathcal{T}_{\alpha}\lsim 0.3$.

\subsection{The Predicted Equivalent Width}
\label{sec:ew1}
While in agreement with the observed LFs, the model described in \S~\ref{sec:lya1} does not reproduce the very large observed equivalent widths. The Ly$\alpha$ luminosity can be rewritten in terms of the star formation rate ($\dot{M}_*$) and EW as $L_{\alpha}=2.0 \times 10^{42}\hs{\rm erg} \hs{\rm s}^{-1}\dot{M}_*(M_{\odot}/{\rm yr})($EW$/160{\rm\hs\AA})$. Here we have used the relation $L_{\alpha}=$EW$\times[\nu_{\alpha}f(\nu_{\alpha})/\lambda_{\alpha}]$, where $f(\nu_{\alpha})$ is the flux density in erg s$^{-1}$ Hz$^{-1}$ at $\nu_{\alpha}$, and where we denoted the Ly$\alpha$ frequency and wavelength by $\nu_{\alpha}$ and $\lambda_{\alpha}$ respectively. Furthermore, we assumed the spectrum to be constant between $1216$ \AA\hs and $1350$ \AA. For $\mathcal{L}_{\alpha,42}=2.0$, we find a best fit model that predicts LAEs to have an observed EW of $\mathcal{T}_{\alpha} \times 160\hs{\rm \AA}\hspace{1mm}\sim 50$ \AA. This value compares unfavorably with the observed sample that includes EWs exceeding $\sim 100$ \AA\hs in $\sim50\%$ of cases for Ly$\alpha$ selected galaxies \citep{Shima06}. 

Thus although our simple model can successfully reproduce the observed Ly$\alpha$ and UV luminosity functions, the model fails to reproduce the observed large EW LAEs. This discrepancy cannot be remedied by changing the intrinsic Ly$\alpha$ emissivity of a given galaxy, $\mathcal{L}_{\alpha}$: increasing $\mathcal{L}_{\alpha}$ would be simply be compensated for by a lower $\mathcal{T}_{\alpha}$ and vice versa. In the next section we discuss a simple modification of this model that aims to alleviate this discrepancy.

\section{The Fluctuating IGM Model}
\label{sec:igm}
The model described in \S~\ref{sec:modelLF} assumed that the Ly$\alpha$ flux of galaxies was subject to uniform attenuation by the IGM. In this section we relax this assumption and investigate the predicted EWs in a more realistic IGM where transmission fluctuates between galaxies. We refer to this model as the 'fluctuating IGM' model. In this model, a larger transmission translates to a larger observed equivalent width. As a result, galaxies with large $\mathcal{T}_{\alpha}$ are more easily detected, and the existence of these galaxies may therefore affect the observed EW-distribution for Ly$\alpha$ selected galaxies, even in cases where they comprise only a small fraction of the intrinsic population. In this section we investigate whether this bias could explain the anomalously large observed EW.
\begin{figure}
\vbox{\centerline{\epsfig{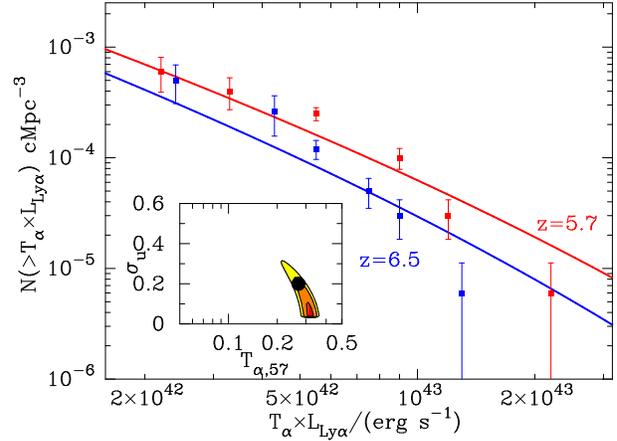}}}
\caption[]{Same as Figure~2. However, instead of assuming a single value of IGM transmission $\mathcal{T}_{\alpha,57}$, we assumed a log-normal distribution of IGM transmission with a mean of $\log \mathcal{T}_{\alpha,57}$ and standard deviation of (in the log) $\sigma_u$ (see text). This reflects the possibility that the IGM transmission fluctuates between galaxies. The inset shows likelihood contours for $(\log \mathcal{T}_{\alpha,57},\sigma_u)$. Increasing $\sigma_u$ flattens the luminosity function (and moves it upward), which is illustrated by the model LFs shown as {\it solid lines}, for which we used $(\mathcal{T}_{\alpha,57},\sigma_u)=(0.27,0.2)$ (shown as the {\it thick black dot} in the inset). The best-fit model to the data has $\sigma_u\sim 0$ (which corresponds to the model shown in Figure~2).}
\label{fig:lya2D}
\end{figure}

We assume a log-normal distribution for $\mathcal{T}_{\alpha,57}$,
\begin{equation}
P(u)du=\frac{1}{\sqrt{2\pi}\sigma_{\rm u}}\times 
\exp\Big{(}\frac{-(u-\langle u\rangle)^2}{2\sigma_{\rm u}^2}\Big{)}du,
\label{eq:lognorm}
\end{equation} where $\langle u\rangle=\log\langle \mathcal{T}_{\alpha,57} \rangle$ is the log (base 10) of the mean transmission and $\sigma_{\rm u}$ is the standard deviation in log-space. Throughout this section we drop the subscript '57'. Eq~(\ref{eq:lognorm}) may be rewritten in the form
\begin{equation}
f(>u)=\frac{1}{2}-\frac{1}{2}{\rm erf}\Big{(}\frac{u-\langle u\rangle}{\sqrt{2}\sigma_u}\Big{)},
\label{eq:lognormcum}
\end{equation} which gives the fraction of LAEs with $\log \mathcal{T}_{\alpha}>u$. The number density of LAEs is then given by 
\begin{equation}
N(>\mathcal{T}_{\alpha}\times L_{\alpha})=\epsilon_{\rm DC}\int_{0}^{\infty}dM\frac{dn}{dM}f(>u(\mathcal{T}_{\alpha} \times L_{\alpha},M))
\label{eq:nlya2} 
\end{equation} where
\begin{equation}
u(\mathcal{T}_{\alpha} \times L_{\alpha},M)=\log\Big{(}\frac{\mathcal{T}_{\alpha}\times L_{\alpha}}{\mathcal{L}_{\alpha}\dot{M}_*}\Big{)}.
\label{eq:ufunc}
\end{equation} Eq~(\ref{eq:nlya2}) differs from Eq~(\ref{eq:form}) in two ways: (1) there is no lower integration limit, and (2) there is an additional term $f(>u)$. These two differences reflect the facts that all masses contribute to the number density of LAEs brighter than $\mathcal{T}_{\alpha}\times L_{\alpha}$, and that lower mass systems require larger transmissions (Eq~\ref{eq:ufunc}) which are less common (Eq~\ref{eq:lognormcum}). In the limit $\sigma_u \rightarrow 0$, the function $f(>u)$ 'jumps' from $0$ to $1$ at $M_{\rm min}$ (Eq~\ref{eq:2}), which corresponds to the original Eq~(\ref{eq:form}).

In this formalism we may also write the number density of LAEs with transmission in the range $u\pm du/2\equiv$ $\log\mathcal{T}_{\alpha}\pm \frac{d\log\mathcal{T}_{\alpha}}{2}$, which is given by
\begin{equation}
N(u)du=P(u)du\int_{\rm M_{\rm min}(u)}^{\infty}dM\frac{dn}{dM}.
\label{eq:ew}
\end{equation} Here $M_{\rm min}(u)$ is the minimum mass of galaxies that can be detected with a transmission in the range $u\pm du/2$ (for $M_{\rm tot} < M_{\rm min}$ the total flux falls below the detection threshold). The number density of LAEs with transmission in the range $u\pm du/2$ may be used to find the number density of LAEs with equivalent widths in the range $\log$ EW$\pm \frac{d\log {\rm EW}}{2}$ via the relation EW$=160\mathcal{T}_{\alpha}$ \AA\hs (for the choice $\mathcal{L}_{\alpha,42}=2.0$, see \S~\ref{sec:lya1}). Eq~(\ref{eq:ew}) shows that the observed equivalent width distribution takes the shape of the original transmission distribution, modulo a boost which increases towards larger EWs.

As in \S~\ref{sec:lya1} we assume the best-fit model parameters for $\epsilon_{\rm DC}$ and $f_*$ derived from the UV-LFs determined in \S~\ref{sec:uv}. We calculate model Ly$\alpha$ LFs on a grid of models in the $(\sigma_u,\langle \mathcal{T}_{\alpha}\rangle)$-plane, and generate likelihoods following the procedure outlined in \S~\ref{sec:lya1}. The results of this calculation are shown in Figure~\ref{fig:lya2D} where we plot the Ly$\alpha$ LFs together with likelihood contours in the $(\sigma_u,\langle \mathcal{T}_{\alpha}\rangle)$-plane (inset). The best-fit models favor no scatter in $\mathcal{T}_{\alpha}$ ($\sigma_u \sim 0$). The reason for this is that for any given model, a scatter in $\mathcal{T}_{\alpha}$ serves to flatten the model LF. However, the observed Ly$\alpha$ LF is quite steep, and as a result the data prefers a model with no scatter. Furthermore, a scatter in $\mathcal{T}_{\alpha}$ results in a model LF that lies above the original (constant transmission) LF at all $\mathcal{T}_{\alpha} \times L_{\alpha}$. This also explains the shape of the contours in the $(\sigma_u,\langle \mathcal{T}_{\alpha}\rangle)$-plane; increasing $\sigma_u$ must be compensated for by lowering $\langle\mathcal{T}_{\alpha}\rangle$). To illustrate the impact of a fluctuating $\mathcal{T}_{\alpha}$, the model LFs shown in Figure~\ref{fig:lya2D} are not those of the best-fit model, but of a model with $(\langle \mathcal{T}_{\alpha}\rangle,\sigma_u)=(0.27,0.2)$.

\begin{figure}
\vbox{\centerline{\epsfig{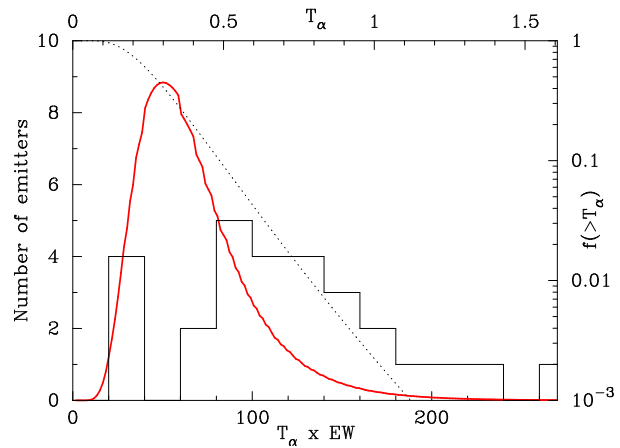}}}
\caption[]{Comparison of the observed equivalent width distribution (EW, {\it histogram}), with the model prediction for a model in which we assumed a log-normal distribution of IGM transmission with $(\mathcal{T}_{\alpha,57},\sigma_u)=(0.27,0.2)$ (see Fig~3). The EW is related to $\mathcal{T}_{\alpha}$ via EW$=160\mathcal{T}_{\alpha}$ \AA. The {\it dotted line} shows the fraction, $f(>\mathcal{T}_{\alpha})$ (shown on the right vertical axis), of galaxies with a transmission greater than $\mathcal{T}_{\alpha}$ (Eq~\ref{eq:lognormcum}). Galaxies with large $\mathcal{T}_{\alpha}$ are more easily detected, hence the large $\mathcal{T}_{\alpha}$ (EW) end is boosted considerably, resulting in closer agreement (but not close enough) to the data.}
\label{fig:ew}
\end{figure}

Figure~\ref{fig:ew} shows the observed EW-distribution ({\it solid line}) at $z=5.7$ associated with the Ly$\alpha$ LFs shown in Figure~\ref{fig:lya2D}. This model may be compared to the observed distribution (shown as the {\it histogram}, data from Shimasaku et al, 2006). The upper horizontal axis shows the transmission corresponding to each equivalent width. The {\it dotted line} shows the distribution of transmission (given by Eq~\ref{eq:lognorm}). Note that the range of transmissions shown in Figure~\ref{fig:ew} extends to $\mathcal{T}_{\alpha}>1$, which of course is not physical. 
The model EW-distribution has a tail towards larger EWs. However, the model distribution peaks at EW$\sim 50$ \AA. As was seen in the constant transmission model, this is clearly inconsistent with the observations, which favor values of EW$\sim 100$ \AA. \citet{Shima06} also show a probability distribution of EW, which is probably closer to the actual distribution. Although this distribution does peak closer to EW$\sim 50$ \AA, it is still significantly broader than that of the model (see \S~\ref{sec:discuss} for more arguments against the fluctuating IGM model). We point out that the model distribution shown in Figure~\ref{fig:ew} is independent of the assumed value of $\mathcal{L}_{\alpha,42}$. 

\section{Ly$\alpha$ Emitters Powered by Very massive Stars.}
\begin{figure*}
\vbox{\centerline{\epsfig{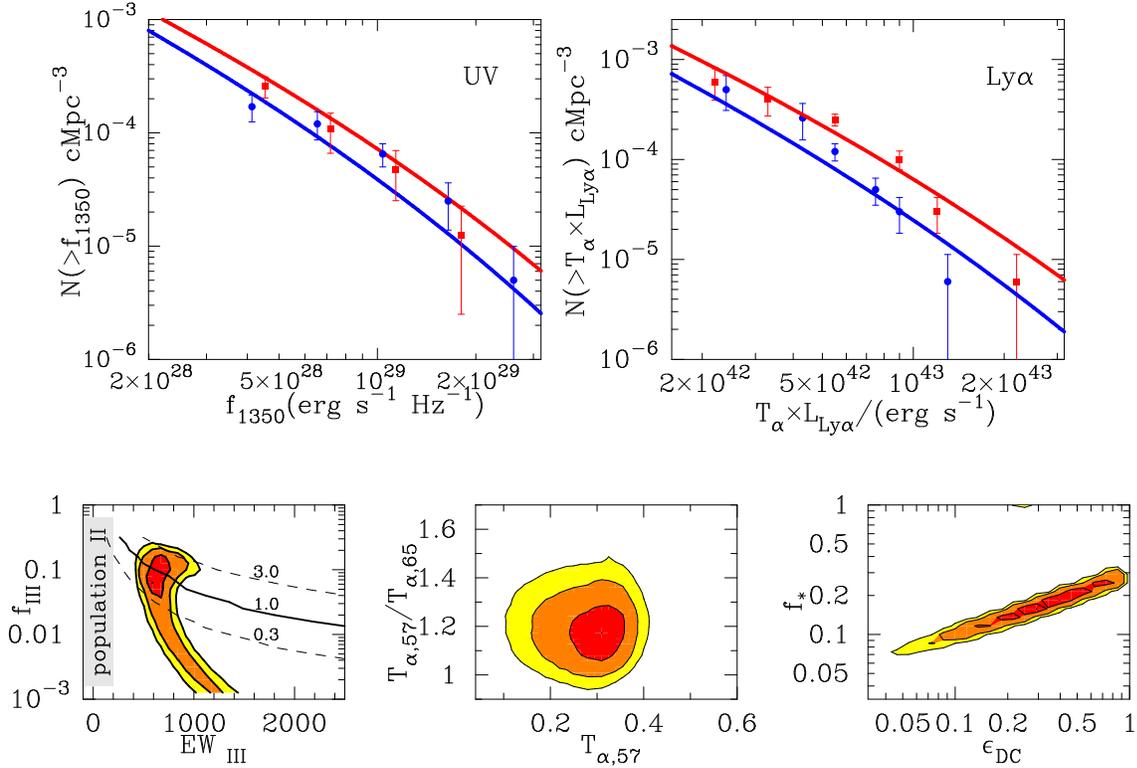}}}
\caption[]{Marginalised constraints on the 6 population-III model parameters, $(\epsilon_{\rm DC},f_{*},\mathcal{T}_{\alpha,57},\mathcal{T}_{\alpha,65},f_{\rm III},{\rm EW}_{\rm III})$, are shown in the lower three panels. In the best-fit population-III model, each galaxy goes through a luminous Ly$\alpha$ phase during which the equivalent width is EW$_{\rm III}=600-800$ \AA\hs for a fraction $f_{\rm III}=0.04-0.1$ of the galaxies life time. Although the bright phase only lasts a few to ten per cent of their life-time, galaxies in the bright phase are more easily detectable, and the number of galaxies detected in Ly$\alpha$ surveys in the bright phase is equal to that detected in the faint phase. This is also demonstrated by the {\it thick solid line} (with label '1.0') in the {\it lower left panel}, which shows the combination of $f_{\rm III}$ and EW$_{\rm III}$ that produces equal numbers of galaxies in the population III and II phase (the {\it dashed lines} are defined similarly, also see Fig~\ref{fig:ewsel}). The best-fit intrinsic equivalent widths are in excess of the maximum allowed for a population-II stellar population having a Salpeter mass function. Therefore, this model requires a burst of very massive star formation lasting no more than a few to ten percent of the galaxies star forming lifetime, and may indicate the presence of population-III star formation in a large number of high-redshift LAEs.}
\label{fig:popIII}
\end{figure*}

\label{sec:popIII}
In \S~\ref{sec:modelLF} we demonstrated that a simple model where $L_\alpha$ and $L_{\rm UV}$ were linearly related to halo mass can reproduce the UV and Ly$\alpha$ LFs, but not the observed EW distribution. In \S~\ref{sec:igm} we showed that this situation is not remedied by a variable IGM transmission, and that favored models have a constant transmission. In this section we discuss an alternate model, which leads to consistency with both the observed Ly$\alpha$ LFs, UV-LFs, and the EW-distribution. In this model, galaxies are assumed to have a bright Ly$\alpha$ phase (hereafter the 'population III'-phase) which lasts a fraction $f_{\rm III}$ of the galaxies' life-time. After this the galaxy's Ly$\alpha$ luminosity drops to the 'normal' value for population II star formation. 

This model may be viewed as an extension of the idea originally described by \citet{MR02}, that large EW LAEs are young galaxies in the early stages of their lives. In this picture, the sudden drop in Ly$\alpha$ luminosity could represent i) a sudden drop in the ionising luminosity when the first O-stars died, or ii) an enhanced dust-opacity after enrichment by the first type-II supernovae. Alternatively, our parametrisation could represent a scenario in which the population III phase ended after the first population III stars enriched the surrounding interstellar gas from which subsequent generations of stars formed. Hence, we refer to this model as the 'population III' model. We will show that to be consistent with the large values of the observed EW, a very massive population of stars is required during the early stages of star formation.

To minimise the number of free parameters we modeled the time dependence of the Ly$\alpha$ EW as a step-function. The number density of LAEs is then given by
\begin{eqnarray}
N(>\mathcal{T}_{\alpha}\times L_{\alpha},z)=f_{\rm III}\times \epsilon_{\rm DC}\int_{M_{\alpha,III}}^{\infty}dM\frac{dn}{dM}(z)+\\ \nonumber
(1-f_{\rm III})\times\epsilon_{\rm DC}\int_{M_{\alpha,II}}^{\infty}dM\frac{dn}{dM}(z).
\label{eq:popIII} 
\end{eqnarray} Here, $M_{\alpha,II}$ is the mass related to $\mathcal{T}_{\alpha}\times L_{\alpha}$ through $\mathcal{T}_{\alpha} \times L_{\alpha}=\mathcal{L}_{\alpha}\times \mathcal{T}_{\alpha}\times\dot{M}_*$, while $M_{\alpha,III}$ is the population III mass, which is calculated with $\mathcal{L}_{\alpha}$ replaced by $\mathcal{L}_{\alpha}=($EW$_{\rm III}/160\hs{\rm\AA})\times 2\times 10^{42}$ erg s$^{-1}$. 

Whereas in previous sections we chose fiducial or best fit parameters for illustration, for the model described in this section we take the most general approach. We fit the model simultaneously to the UV-LF and Ly$\alpha$ LFs, as well as to the observed EW-distribution of Ly$\alpha$ selected galaxies. This model predicts two observed equivalent widths ($\mathcal{T}_{\alpha}\times$EW$_{\rm III}$ and $\mathcal{T}_{\alpha}\times$EW$_{\rm II}$) in various abundances. The associated mean and variance from the model are compared to the observed EW-distribution, which has a mean of $\langle$EW$\rangle=120\pm25$ \AA, and a standard deviation of $\sigma_{\rm EW}=50 \pm 10$ \AA.

Our model has 6 free parameters $(\epsilon_{\rm DC},f_{*},\mathcal{T}_{\alpha,57},\mathcal{T}_{\alpha,65},f_{\rm III},{\rm EW}_{\rm III})$. We produce likelihoods for each parameter by marginalising over the others in this space. The lower set of panels in Figure~\ref{fig:popIII} show likelihood contours for our model parameters at $64\%$, $26\%$ and $10\%$ of the peak likelihood. The best-fit models have EW$_{\rm III}\sim 600-800$ \AA\hspace{1mm} and $f_{\rm III}=0.04-0.1$ which corresponds to a physical timescale for the population-III phase of $f_{\rm III}\times \epsilon_{\rm DC}\times t_{\rm hub}\sim 4-50$ Myr (for $0.1  \lsim \epsilon_{\rm DC} \lsim 0.5$). The model Ly$\alpha$ luminosity functions at z=5.7 and z=6.5 described by $(\epsilon_{\rm DC},f_{*},\mathcal{T}_{\alpha,57},\mathcal{T}_{\alpha,65},f_{\rm III},{\rm EW}_{\rm III})=$ $(0.2,0.14,0.22,0.19,0.08,650$ \AA$)$ are shown as solid lines and provide good fits to the data. The model produces two observed EWs, namely ${\mathcal T}_\alpha \times$EW$_{\rm II}=35$ \AA\hspace{1mm} and ${\mathcal T}_\alpha\times $ EW$_{\rm III}=143$ \AA\hspace{1mm}. It is worth emphasising that the emitted EW of the bright phase depends on the choice $\mathcal{L}_{\alpha,42}$ via EW$_{\rm III}=650(\mathcal{L}_{\alpha,42}/2.0)$ \AA. Hence, a lower/higher value of $\mathcal{L}_{\alpha,42}$ would decrease/increase the intrinsic brightness of the 'population III' phase. Note that $\mathcal{L}_{\alpha,42}=1.0$ when LAEs formed out of gas of solar metallicity, which is unreasonable given the universe was only $\sim 1$ Gyr old at $z=6$. Furthermore, $\mathcal{L}_{\alpha,42}=1$ would have yielded a best-fit $\mathcal{T}_{\alpha,57}=0.5$, which is well outside the range calculated by \citet{IGM}. We there conclude that $\mathcal{L}_{\alpha,42}$ is in excess of unity.
 
In performing fits we have fixed the value of EW$_{\rm II}$ to correspond to a standard stellar population, and then explored the possibility that there might be a second phase of SF producing a larger EW. Our modeling finds strong statistical evidence for this early phase and rules out the null-hypothesis that properties can be described by population-II stars alone at high confidence ({\it grey region} in the {\it lower left panel} inset of Fig~\ref{fig:popIII}). Despite the fact that the best fit model has a bright phase which lasts only a few per cent of the total star formation lifetime, the two populations of LAEs are similarly abundant in model realisations of the observed sample in Ly$\alpha$ selected galaxies (see \S~\ref{sec:uvew} for a more detailed comparison to the observed EW distributions). This is shown in the {\it lower left panel} in Figure~\ref{fig:popIII} in which the {\it solid line} (with label '1.0') shows the combination of $f_{\rm III}$ and EW$_{\rm III}$ for which the observed number of galaxies in the population III phase ($N_{\rm III}$) equals that in the population II phase ($N_{\rm II}$). The {\it dashed lines} show the cases $N_{\rm III}/N_{\rm II}=0.3$ and  $N_{\rm III}/N_{\rm II}=3.0$. The duration of the bright Ly$\alpha$ phase meets theoretical expectations for a burst of star formation, while the large EW requires a very massive stellar population \citep[e.g][]{S03,T03}. In summary, in order to reproduce both the UV and Ly$\alpha$ LFs, and the observed population of large EW galaxies, we require a burst of very massive star formation lasting $\lsim 10$ per cent of the galaxies lifetime.

\subsection{EW Distribution of UV and Ly$\alpha$ Selected Galaxies}
\label{sec:uvew}
\citet{Stanway07} show that 11 out of 14 LAE candidates among i-drop galaxies in the Hubble Ultra Deep Field have EW$<100$ \AA. If galaxies are included for which only upper or lower limits on the EW are available, then this fraction becomes 21 out of 26. Thus the distribution of EWs for i-drop selected galaxies differs strongly from the EW-distribution observed by \citet{Shima06}. We next describe why this strong dependence of the observed Ly$\alpha$ EW distribution on the precise galaxy selection criteria arises naturally in our population III model. 

We first assume that the Ly$\alpha$ selected and UV-selected galaxies were drawn from the same population (this assumption is discussed further in \S~\ref{sec:q}). In our model a galaxy that is selected based on it's rest-frame UV-continuum emission has a probability $f_{\rm III}$  of being observed in the Ly$\alpha$ bright phase, while the probability of finding a galaxies in the Ly$\alpha$ faint phase is $1-f_{\rm III}$. In \S~\ref{sec:popIII} we found $f_{\rm III}\sim 0.1$, hence an i-drop galaxy is $\sim 10$ times more likely to have a low than a high observed EW. If we denote the number of galaxies with EW$>100$ \AA\hs by $N_{\rm III}$, and the number of galaxies with EW$<100$ \AA\hs by $N_{\rm II}$, then the model predicts $N_{\rm III}/N_{\rm II}=$$f_{\rm III}/(1-f_{\rm III})\sim 0.1$, while the observed fraction including the galaxies for which the EW is known as upper or lower limit is $N_{\rm III}/N_{\rm II}=0.19\pm 0.05$. Therefore the qualitative difference in observed Ly$\alpha$ EW distribution among i-drop galaxies in the HUDF and among Ly$\alpha$ selected galaxies follows naturally from our two-phase star formation model. Note that our model predicts population III star formation to be observed in $f_{\rm III}/(1-f_{\rm III})\sim 10\%$ of the $z=6.0$ LBG population.

\begin{figure}
\vbox{\centerline{\epsfig{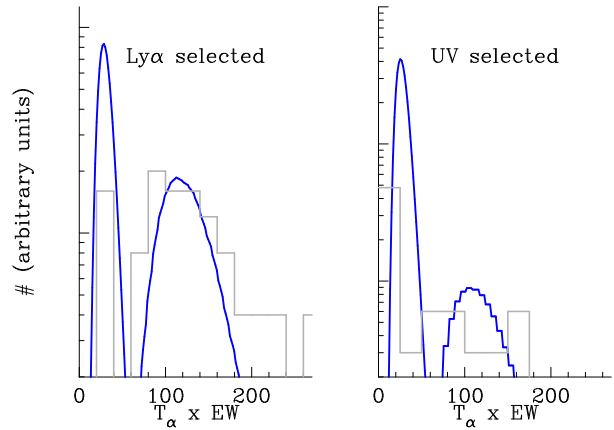}}}
\caption[]{Comparison of the predicted EW distribution for UV and Ly$\alpha$ selected galaxies. The best-fit population-III model (see \S~\ref{sec:popIII}) was used. In order to get a finite range of observed EWs (instead of only two values at $\mathcal{T}_{\alpha}\times$EW$_{\rm II}$ and $\mathcal{T}_{\alpha}\times$EW$_{\rm III}$, we assumed the IGM transmission to fluctuate. The units on the vertical axis are arbitrary. The figure shows that in our population III model, the Ly$\alpha$ selected sample contains a larger relative fraction of large EW LAEs than the UV-selected (i-drop) sample, which is qualitatively 
in good agreement with the observations (shown by the {\it histograms}).}
\label{fig:ewsel}
\end{figure}

The dependence of the observed EW distribution on the selection criteria used to construct the sample of galaxies is illustrated in Figure~\ref{fig:ewsel}. To construct this figure, we have taken the best-fit population III model of \S~\ref{sec:popIII}. For the purpose of presentation, we let the IGM fluctuate according to the prescription of \S~\ref{sec:igm} with $\sigma_u=0.1$, so that the model predicts a finite range of EWs in each phase.
 The {\it left and right panels} show the predicted EW distribution for Ly$\alpha$ selected ({\it left panel}) and UV-selected ({\it right panel}) galaxies as the {\it solid lines}, respectively. For a UV-selected galaxy the probability of being in the bright phase and having an observed EW in the range EW$_{\rm III}\times (\mathcal{T}_{\alpha}\pm d\mathcal{T}_{\alpha}/2)$ is $f_{\rm III}P(\mathcal{T}_{\alpha})d\mathcal{T}_{\alpha}$. Here $P(\mathcal{T}_{\alpha})d\mathcal{T}_{\alpha}$ is the probability that the IGM transmission is in the range $\mathcal{T}_{\alpha}\pm d\mathcal{T}_{\alpha}/2$, which is derived from Eq~\ref{eq:lognorm}. The units on the vertical axis are arbitrary, and chosen to illustrate the different predicted and observed Ly$\alpha$ EW distribtions for the two samples at large EWs. The observed distributions for Ly$\alpha$ and UV selected galaxies, shown as {\it histograms}, are taken from \citet{Shima06} and \citet{Stanway07}, respectively. Figure~\ref{fig:ewsel} clearly shows that both the predicted and observed Ly$\alpha$ selected samples contain significantly more large EW LAEs than the UV-selected sample. Our model naturally explains the qualitative shape of these distributions and their differences. 

Before proceeding we mention a caveat to the distributions shown in Figure~\ref{fig:ewsel}. In our model all galaxies have an EW of $\mathcal{T}_{\alpha}\times$EW$_{\rm II}\sim 25-35$ \AA\hs during the population II phase, while in contrast \citet{Stanway07} do not detect 10 out of 26 LBGs, which implies that $\sim 40\%$ of LBGs have an EW$\lsim 6$ \AA. Thus there is a descrepancy between our model and the observations with respect to the value of EW in the population II phase. The resolution of this discrepancy lies in the fact that the very low EW emitters are drawn from the UV (i-drop) sample and not the Ly$\alpha$ selected sample our model was set up to describe. This issue is discussed in more detail in \S~\ref{sec:q}.

An EW distribution of dropout sources was also presented by \citet{D07}. These authors performed an analysis similar to \citet{Stanway07} and found 1 LAE with EW$=150$ \AA\hs among 22 candidate z=6.0 LBGs. When interpreted in reference to our model, this translates to $N_{\rm III}/N_{\rm II}\sim 5\%$, which is consistent with the model predictions. Therefore, when interpreted in light of a two-phase star formation history and different selection methods, the EW distribution observed by \citet{D07} is consistent with that found by \citet{Shima06}.

If population III star formation does provide the explanation for the very large EW Ly$\alpha$ emitters, then we would expect the large EW emitters to become less common with time as the mean metallicity of the Universe increased. To test this idea, we can compare the EW-distribution at $z=5.7$ with the results at lower redshift from \citet{Shapley03} who found that $\lsim 0.5\%$ of $z=3$ LBGs have Ly$\alpha$ EW$\geq150$ \AA\hs, and that $\lsim 2\%$ of $z=3$ LBGs to have Ly$\alpha$ EW$\geq 100$ \AA. \citet{D07} argue that the fraction of large EW Ly$\alpha$ lines at $z=6$ is consistent with that observed at $z=3$ \citep{Shapley03}. However, if the EW distribution did not evolve with redshift, then the probability that a sample of 22 LBGs will contain at least 1 LAE with EW$\gsim 150$ \AA\hs is $\lsim 10\%$. Thus the hypothesis that the observed EW distribution remains constant is ruled out at the $\sim 90\%$ level.  On the other hand, in a similar analysis \citet{Stanway07} found 5 out of 26 LBGs to have an EW$\gsim 100$ \AA. If the EW distribution did not evolve with redshift, then the probability of finding 5 EW$\gsim 100$ \AA\hs in this sample is only $\sim 10^{-4}$. Furthermore, \citet{Nagao07} recently found at least 5 LAEs with EW$>100$ \AA\hs at $6.0\lsim z\lsim 6.5$, and conclude that 8\% of i'-drop galaxies in the Subaru Deep Field have EW$>100$ \AA, which is significantly larger than the fraction of large EW LBGs at $z=3$. Therefore, the observed EW distribution of LBGs at $z=6$ is skewed more toward large EWs than at $z=3$. The strength of this result is increased by the fact that the IGM is more opaque to Ly$\alpha$ photons at $z=6$ than at $z=3$. Thus we conclude that the intrinsic EW distribution must have evolved with redshift.

\section{Discussion}
\label{sec:discuss}

\subsection{Comparison with Population Synthesis Models}
\label{sec:syn}
Population synthesis models have suggested that the broad band colors of observed LAEs are best described with young stellar populations \citep{Gawiser06,Pirzkal07,Finkelstein07}. \citet{Lai07} found the stellar populations in three LAEs to be $5-100$ Myr old, and possibly as old as $700$ Myr (where the precise age upper limit depends on the assumed star formation history of the galaxies). However as was argued by \citet{Pirzkal07}, since these galaxies were selected based on their detection in IRAC, a selection bias towards older stellar populations may exist \citep[also see][]{Lai07}. Furthermore, \citet{Finkelstein07} found that LAEs with EW$>110$ \AA\hs have ages $\lsim$4 Myr, while LAEs with EW$<40$ \AA\hs have ages between 20-400 Myr. This latter result in particular agrees well with our population III model. On the other hand, in a fluctuating IGM model for example, the EW of LAEs should be uncorrelated with age. 

In models presented in this paper, on average $f_*\sim 0.15$ of all baryons are converted into stars within halos of mass $M_{\rm tot}\sim 10^{10}-10^{11}M_{\odot}$, yielding stellar masses in the range $M_*=10^8-10^9M_{\odot}$ \citep{LF,Stark07}. This compares unfavorably with the typical stellar masses found observationally in LAEs which can be as low as $M_*=10^6-10^7M_{\odot}$ \citep{Finkelstein07,Pirzkal07}. However, the lowest stellar masses are found (naturally) for the younger galaxies. Indeed, the LAEs with the oldest stellar populations can have stellar masses as large as $10^{10}M_{\odot}$. Thus, we do not find the derived stellar masses in LAEs to be at odds with the results of this paper.  
If significant very massive (or population III) star formation indeed occurred in high redshift LAEs, then one may expect these stars to reveal themselves in unusual broad-band colors \citep[e.g.][]{Stanway05}. However, \citet{T03} have shown that the most distinctive feature in the spectrum of population III stars is the number of H and He ionising photons \citep[also see][]{Bromm01}. Since these are (mostly) absorbed in the IGM, the broad band spectrum of population III stars is in practice difficult to distinguish from a normal stellar population \citep{T03}, especially when nebular continuum emission is taken into account \citep[][see their Fig~1]{SP05}. Hence, population III stars would not necessarily be accompanied by unusually blue broad band colors. 
 
\subsection{Alternative Explanations for Large EW LAEs}
\label{sec:fluc}
We have shown that a simple model in which high-redshift galaxies go through a population-III phase lasting $\lsim15$ Myr can simultaneously explain the observed Ly$\alpha$ LFs at $z=5.7$ and $z=6.5$ \citep{Ka06}, and the observed EW-distribution of Ly$\alpha$ selected galaxies at $z=5.7$ \citep{Shima06}. In addition, this model predicts the much lower EWs found in the population of UV selected galaxies (Stanway et al, 2007, see \S~\ref{sec:uvew}). Moreover the constraints on the population-III model parameters such as the duration and the equivalent width of the bright phase are physically plausible, and consistent with existing population synthesis work (see \S~\ref{sec:syn}).

Are there other interpretations of the large observed EWs? One possibility was discussed in \S~\ref{sec:igm}, where we showed that the simple model in which the IGM transmission fluctuates between galaxies reproduces the LFs, but fails to simultaneously reproduce the Ly$\alpha$ LFs and the observed EW-distribution. In addition, this model fails to reproduce other observations. \citet{IGM} calculated the impact of the high-redshift reionised IGM on Ly$\alpha$ emission lines and found the range of plausible transmissions to lie in the range $0.1<\mathcal{T}_{\alpha}<0.3$. This work showed that it is possible to boost the transmission to (much) larger values but not without increasing the observed width of the Ly$\alpha$ line. Absorption in the IGM typically erases all flux blueward of the Ly$\alpha$ resonance, and when infall is accounted for, part of the Ly$\alpha$ redward of the Ly$\alpha$ resonance as well. This implies that Ly$\alpha$ lines that are affected by absorption in the IGM are systematically narrower than they would have been if no absorption in the IGM had taken place. It follows that in the fluctuating IGM model, Ly$\alpha$ EW should be strongly correlated with the observed Ly$\alpha$ line width (or FHWM). This correlation is not observed. In fact, observations suggest that an anti-correlation exists between EW and FWHM \citep[][]{Shima06,Tapken07}. This anti-correlation provides strong evidence against the anomalously large EWs being produced by a fluctuating IGM transmission.

A second possibility is the presence of galaxies with strong superwinds. The models of \citet{IGM} did not study the impact of superwinds on the Ly$\alpha$ line profile. The presence of superwinds can cause the Ly$\alpha$ line to emerge with a systematic redshift relative to the Ly$\alpha$ resonance through back scattering of Ly$\alpha$ photons off the far side of the shell that surrounds the galaxy \citep{Ahn03,Ahn04,Hansen06,V06}. However, superwinds tend not only to redshift the Ly$\alpha$ line, they also make the Ly$\alpha$ line appear broader than when this scattering does not occur. As in the case of the fluctuating IGM model, this results in a predicted correlation between EW and FWHM, which is not observed. Furthermore in wind-models, the overall redshift of the Ly$\alpha$ line, and thus $\mathcal{T}_{\alpha}$, increases with wind velocity, $v_w$. This predicts that EW increases with wind velocity. However, observations of $z=3$ LBGs by \citet{Shapley03} show that EW correlates with $v_w^{-1}$ \citep{Ferrara06}. We therefore conclude that the large EW in LAEs cannot be produced by superwind galaxies. 

 A third possibility might be that within the Ly$\alpha$ emitting galaxy, cold, dusty clouds lie embedded in a hot inter-cloud medium of negligible Ly$\alpha$ opacity. Under such conditions, the continuum photons can suffer more attenuation than Ly$\alpha$ photons which bounce from cloud to cloud and mainly propagate through the hot, transparent inter-cloud medium \citep{Neufeld91,Hansen06}. This attenuation of continuum leads to a large EW. We point out that in this scenario, large EW LAEs are not intrinsically brighter in Ly$\alpha$. At fixed Ly$\alpha$ flux, one is therefore equally likely to detect a low EW LAE. In other words, to produce the observed EW distribution one requires preferential destruction of continuum flux by dust in $\sim 50\%$ of the galaxies. Currently there is no evidence that this mechanism is at work even in one galaxy. Furthermore, the rest-frame UV colors of galaxies in the Hubble Ultra Deep Field imply that dust in high-redshift galaxies suppresses the continuum flux by only a factor of $\sim 1.4$ \citep{Bouwens06}. The maximum boost of the EW in a multi-phase ISM is therefore $1.4$, which is not nearly enough to produce intrinsic equivalent widths of EW$\sim 600-800$ \AA. In summary, the only model able to simultaneously explain all observations calls for a short burst of very massive star formation.

\subsection{Comparison with the LBG Population}
\label{sec:q}
\begin{figure}
\vbox{\centerline{\epsfig{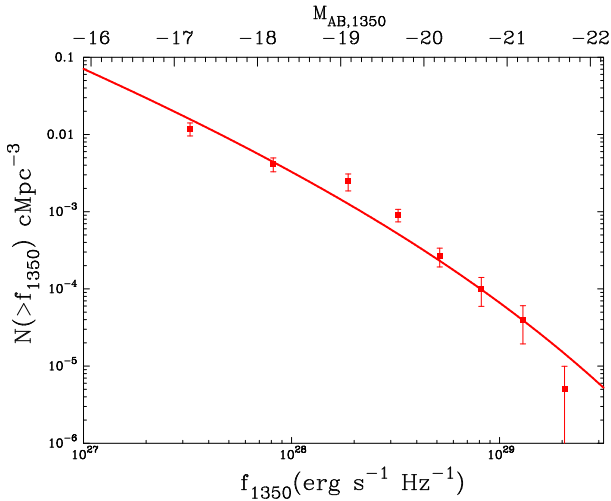}}}
\caption[]{Comparison of the best-fit population III model (shown in Figure~\ref{fig:popIII}) with the UV-LF constructed by \citet{Bouwens06} using $\sim 300$ LBGs in the Hubble Deep Fields. This good agreement is likely a coincidence since in our model all LBGs have an observed EW$\gsim 30$ \AA, while the observations show that $\sim 40\%$ of all LBGs are not detected in Ly$\alpha$ (EW$\lsim 6$ \AA). This overlap could possibly be (partly) due to a lower dust content of LAEs relative to their Ly$\alpha$ quiet counterparts (see text).}
\label{fig:bouwens}
\end{figure}
In \S~\ref{sec:uvew} we have shown that the observed EW distributions of Ly$\alpha$ selected and i-drop galaxies and their differences can be reproduced qualitatively with our population III model. However, in our model all high-redshift galaxies have an observed EW of at least $\mathcal{T}_{\alpha}\times$EW$_{\rm II}\sim 30$ \AA, whereas many i-drop galaxies are not detected in Ly$\alpha$. \citet{Ka06} show that the UV-LFs of LAEs at $z=6.5$ and $z=5.7$ overlap with that constructed by \citet{Bouwens06} from a sample of $\sim 300$ $z=6$ LBGs discovered in the Hubble Deep Fields. Naively, this overlap implies that LBGs and LAEs are the same population and therefore that all LBGs should be detected by Ly$\alpha$ surveys. Since Ly$\alpha$ surveys only detect galaxies with EW$\gsim 20$ \AA, this suggests all LBGs should have a Ly$\alpha$ EW \gsim 20 \AA\hs contrary to observation. To illustrate this point further, we have taken the best-fit population III model shown in Figure~\ref{fig:popIII} and compared the model predictions for the rest-frame UV-LF with that of \citet{Bouwens06} in Figure~\ref{fig:bouwens}. Clearly, our best-fit population III model fits the data well. However, \citet{Stanway07} found $\sim 40\%$ of i-drop galaxies in the HUDF to have an observed EW$\lsim 6$ \AA, and a similar result was presented by \citet{D07}.

Two effects may help reconcile these two apparently conflicting sets of observations: (i) \citet{D07} found Ly$\alpha$ emitting LBGs to be systematically smaller. That is, for a fixed angular size, the $z_{850}$-band flux of LAEs is systematically higher with $\Delta z_{850}\sim -1$. If we assume that the angular scale of a galaxy is determined by the mass of its host halo, then this implies that for a fixed mass the $z_{850}$-band flux of LAEs is systematically higher, and (ii) only a fraction of LBGS are LAEs. The drop-out technique used to select high redshift galaxies is known to introduce a bias against strong LAEs, as a strong Ly$\alpha$ line can affect the broad band colors of high-redshift galaxies. This may cause $\sim 10-46\%$ of large EW LAEs to be missed using the i-drop technique \citep{D07}.

If only a fraction $f_{\alpha}$ of all LBGs are detected in Ly$\alpha$, then effect (i) would explain why the UV-LF of LAEs lies less than a factor of $1/f_{\alpha}$ below the observed UV-LF of the general population of LBGs. This is because a more abundant lower mass halo is required to produce the same UV-flux in LAEs, which would shift the LF upwards. In addition, effect (ii) may reduce this difference even further. It follows that these two effects combined may cause the LFs to overlap. Thus the overlap of the UV and Ly$\alpha$ selected UV-LFs appears to be a coincidence, and not evidence of their being the same population of galaxies. This implies that our model is valid for Ly$\alpha$ selected galaxies, but not the high-redshift population as a whole and explains the lack of very low EWs in UV selected samples discussed in \S~\ref{sec:uvew}.

 The reason why LAEs may be brighter in the UV for a fixed halo mass is unclear. It is possibly related to dust content. \citet{Bouwens06} found the average amount of UV exctinction to be 0.4 mag in the total sample of $z=6$ LBGs. This value is close to the average excess $z_{850}$-band flux detected from LAEs for a given angular scale \citep[][]{D07}. If LAEs contain less (or no) dust, then this would explain why they are brighter in the UV and thus why they appear more compact. The possibility that 'Ly$\alpha$ quiet' LBG contain more dust than their Ly$\alpha$ emitting counterparts is not very surprising, as a low dust abundance has the potential to eliminate the Ly$\alpha$ line. Thus LAEs could be high redshift galaxies with a lower dust content.

\citet{Shima06} and \citet{Ando06} found that luminous LBGs, $M_{\rm UV}\lsim -21.0$, typically do not contain large EW Ly$\alpha$ emission lines. This deficiency of large EW LAEs among UV-bright sources is not expected in our model, and may reflect that UV-bright sources are more massive, mature, galaxies that cannot go through a population-III phase anymore. It should be pointed out though that the absence of large EW LAEs among galaxies with $M_{\rm UV}\lsim -21.0$ in the survey of \citet{Shima06} is consistent with our model: The observed number density of sources with $M_{\rm UV}\lsim -21.0$ is $\sim 5\times 10^{-5}$ cMpc$^{-3}$ (see Fig~\ref{fig:uv}). In our best-fit pop-III model (shown in Fig~\ref{fig:popIII}), a fraction $f_{\rm III}\sim 0.08$ of these galaxies would be in the bright phase. This translates to a number density of large EW LAEs of $\sim 4\times 10^{-6}$ cMpc$^{-3}$. Given the survey volume of $\sim 2\times 10^5$ cMpc$^3$, the expected number of large EW LAEs with $M_{\rm UV}\lsim -21.0$ is $\sim 0.8$, and the absence of large EW LAE among UV bright sources is thus not surprising.

\subsection{Clustering Properties of the LAEs}

In our model large EW LAEs are less massive by a factor of EW$_{\rm III}/$EW$_{\rm II}\sim 4$ at fixed Ly$\alpha$ luminosity. Since clustering of dark matter halos increases with mass, it follows that our model predicts large EW LAEs to be clustered less than their low EW counterparts (at a fixed Ly$\alpha$ luminosity). The clustering of LAEs is typically quantified by their angular correlation function (ACF), $w(\theta)$, which gives the excess (over random) probability of finding a pair of LAEs separated by an angle $\theta$ on the sky. The ACF depends on the square of the bias parameter ($w(\theta)\propto b^2(m)$), which for galaxies in the population II phase is $\sim 1.24-1.4$ times larger than for galaxies in the population III phase, for the mass range of interest. This implies that the clustering of low EW LAEs at fixed Ly$\alpha$ luminosity is enhanced by a factor of $\sim 1.5-2.0$. Existing determinations of the ACF of LAEs by \citet{Shima06} and \citet{Ka06} are still too uncertain to test this prediction.

\section{Conclusions}
\label{sec:conclusion}
Observations of high redshift Ly$\alpha$ emitting galaxies (LAEs) have shown the typical equivalent width (EW) of the Ly$\alpha$ line to increase dramatically with redshift, with a significant fraction of the galaxies lying at $z\geq 5.7$ having an EW$\gsim 100$~\AA. Recent calculations by \citet{IGM} show that the IGM at $z\geq 4.5$ transmits only $10-30\%$ of the Ly$\alpha$ photons emitted by galaxies. In this paper we have investigated the transmission using a model that reproduces the observed Ly$\alpha$ and UV LFs. This model results in an empirically determined transmission of $\mathcal{T}_{\alpha}\sim 0.30(\mathcal{L}_{\alpha,42}/2.0)^{-1}$, where $\mathcal{L}_{\alpha,42}$ denotes the Ly$\alpha$ luminosity per unit star formation rate (in $M_{\odot}$ yr$^{-1}$) $\mathcal{L}_{\alpha}$ in units of $10^{42}\hs{\rm erg} \hs{\rm s}^{-1}$ (\S~\ref{sec:modelLF}). This value is in good agreement with earlier theoretical results. 

If only $\sim 30\%$ of all Ly$\alpha$ that was emitted by high redshift galaxies reaches the observer, then this impies that the intrinsic EWs are systematically (much) larger than observed in many cases. To investigate the origin of these very high EWs, we have developed semi-analytic models for the Ly$\alpha$ and UV luminosity functions and the distribution of equivalent widths. In this model Ly$\alpha$ emitters undergo a burst of very massive star formation that results in a large intrinsic EW, followed by a phase of population-II star formation that produces a lower EW\footnote{Technically, the model discussed in \S~\ref{sec:popIII} only specifies that galaxies go through a 'population-III phase for a fraction $f_{\rm III}\sim 0.1$ of their lifetimes. Our model does not specify when this population-III phase occurs. Hypothetically, the population III phase could occur at an arbitrary moment in the galaxies' lifetime when it is triggered by a merger of a regular star forming galaxy and a dark matter halo containing gas of primordial composition. Note however that such a model would probably have difficulties explaining the apparent observed correlation between Ly$\alpha$ EW and the age of a stellar population (\S~\ref{sec:syn}).}. This model is referred to as the 'population III model' and is an extension of the idea originally described by \citet{MR02}, who proposed large EW Ly$\alpha$ emitters to be young galaxies. 

The population III model in which the Ly$\alpha$ equivalent width is EW$_{\rm III}\sim 650(\mathcal{L}_{\alpha,42}/2.0)$ \AA\hs for $\lsim 50$ Myr, is able to simultaneously describe the following eight properties of the high redshift galaxy population: i-iv) the UV and Ly$\alpha$ luminosity functions of LAEs at z=5.7 and 6.5, v-vi) the mean and variance of the EW distribution of Ly$\alpha$ selected galaxies at z=5.7, vii) the EW distribution of UV- selected galaxies at z$\sim$6 (\S~\ref{sec:popIII}), and viii) the observed correlation of stellar age and mass with EW (\S~\ref{sec:syn}). Our modeling suggests that the anomalously large intrinsic equivalent widths observed in about half of the high redshift Ly$\alpha$ emitters require a burst of very massive star formation lasting no more than a few to ten percent of the galaxies star forming lifetime. This very massive star formation may indicate the presence of population-III star formation in a large number of high-redshift LAEs. The model parameters for the best-fit model are physically plausible where not previously known (e.g. those related to the efficiency and duration of star formation), and agree with estimates where those have been calculated directly (e.g the IGM transmission, EW$_{\rm III}$, and $f_{\rm III}$). 

In addition, we argued that the observed overlap of the UV-LFs of LAEs with that of z$\sim 6$ LBGs appears to be at odds the observed Ly$\alpha$ detection rate in high-redshift LBGs, suggesting that LAEs and LBGs are not the same population. A lower dust content of LAEs relative to their `Ly$\alpha$ quiet' counterparts would partly remedy this discrepancy, and could also explain why LAEs appear to be typically more compact (\S~\ref{sec:q}).

Semi-analytic modeling of the coupled reionisation and star formation histories of the universe suggests that population III star formation could still occur after the bulk of reionisation had been completed \citep{Sc03,S06,WC}. The observation of anomalously large EWs in Ly$\alpha$ emitting galaxies at high redshift may therefore provide observational evidence for such a scenario. In the future, the He 1640 \AA\hspace{1mm} may be used as a complementary probe \citep[e.g][]{T01,T03}. The EW of this line is smaller by a factor of $\gsim 20$ for population III \citep{S03}. However, the He 1640\AA\hspace{1mm} will not be subject to a small transmission of $\sim 10-30\%$, making it accessible to the next generation of space telescopes. On the other hand, it may also be possible to observe the He 1640 \AA\hspace{1mm} in a composite spectra of $z=6$ LBGs. Indeed, the He 1640 \AA\hspace{1mm} line has already been observed in the composite spectrum of z$=3$ LBGs \citep{Shapley03}, which led \citet{JH06} to argue for population III star formation at redshifts as low as $z=3-4$. If population III star formation was more widespread at higher redshifts, as predicted by our model, then the composite spectrum of LBGs at higher redshifts should exhibit an increasingly prominent He 1640 \AA\hspace{1mm} line. In particular, this line should be most prominent in the subset of LBGs that have large EW Ly$\alpha$ emission lines.

{\bf Acknowledgments} JSBW and MD thank the Australian Research Counsel
for support. We thank Avi Loeb for useful discussions, and an anonymous referee for a helpful report that improved the content of this paper.

\label{lastpage}

\begin{thebibliography}{14}
\expandafter\ifx\csname natexlab\endcsname\relax\def\natexlab#1{#1}\fi

\bibitem[Ahn et al.(2003)]{Ahn03} Ahn, S.-H., Lee, H.-W., \& 
Lee, H.~M.\ 2003, \mnras, 340, 863 

\bibitem[Ahn(2004)]{Ahn04} Ahn, S.-H.\ 2004, \apjl, 601, L25 

\bibitem[Ando et al.(2006)]{Ando06} Ando, M., Ohta, K., Iwata, 
I., Akiyama, M., Aoki, K., \& Tamura, N.\ 2006, \apjl, 645, L9 

\bibitem[Barkana(2004)]{Barkana-Infall} Barkana, R.\ 2004, \mnras, 
347, 59 
\bibitem[Bouwens et al.(2006)]{Bouwens06} Bouwens, R.~J., 
Illingworth, G.~D., Blakeslee, J.~P., \& Franx, M.\ 2006, \apj, 653, 53 

\bibitem[Bromm et al.(2001)]{Bromm01} Bromm, V., Kudritzki, 
R.~P., \& Loeb, A.\ 2001, \apj, 552, 464 

\bibitem[Charlot \& Fall(1993)]{CF93} Charlot, S., \& Fall, 
S.~M.\ 1993, \apj, 415, 580 

\bibitem[Dawson et al.(2004)]{Dawson04} Dawson, S., et al.\ 
2004, \apj, 617, 707 

\bibitem[Dijkstra et al.(2007)]{IGM} Dijkstra, M., Lidz, 
A., \& Wyithe, J.~S.~B.\ 2007, \mnras, 377, 1175 

\bibitem[Dijkstra et al.(2007b)]{LF} Dijkstra, M., Wyithe, 
J.S.B., \& Haiman, Z.,\ 2007b, \mnras \space{1mm} in press, astroph/0611195

\bibitem[Dow-Hygelund et al.(2007)]{D07} Dow-Hygelund, 
C.~C., et al.\ 2007, \apj, 660, 47 

\bibitem[Ferrara \& Ricotti(2006)]{Ferrara06} Ferrara, A., \& 
Ricotti, M.\ 2006, \mnras, 373, 571 

\bibitem[Finkelstein et al.(2007)]{Finkelstein07} Finkelstein, S.~L., 
Rhoads, J.~E., Malhotra, S., Pirzkal, N., \& Wang, J.\ 2007, \apj, 660, 
1023 

\bibitem[Gawiser et al.(2006)]{Gawiser06} Gawiser, E., et al.\ 
2006, \apjl, 642, L13 

\bibitem[Hansen \& Oh(2006)]{Hansen06} Hansen, M., \& Oh, S.~P.
2006, \mnras, 367, 979 

\bibitem[Hu \& McMahon(1996)]{Hu96} Hu, E.~M., \& McMahon, 
R.~G.\ 1996, \nat, 382, 231 

\bibitem[Hu et al.(2002)]{Hu02} Hu, E.~M., Cowie, L.~L., 
McMahon, R.~G., Capak, P., Iwamuro, F., Kneib, J.-P., Maihara, T., \& 
Motohara, K.\ 2002, \apjl, 568, L75 

\bibitem[Hu et al.(2004)]{Hu04} Hu, E.~M., Cowie, L.~L., 
Capak, P., McMahon, R.~G., Hayashino, T., \& Komiyama, Y.\ 2004, \aj, 127, 
563 

\bibitem[Iye et al.(2006)]{Iye06} Iye, M., et al.\ 2006, 
\nat, 443, 186 

\bibitem[Jimenez \& Haiman(2006)]{JH06} Jimenez, R., \& 
Haiman, Z.\ 2006, \nat, 440, 501 

\bibitem[Kashikawa et al.(2006)]{Ka06} Kashikawa, N., et 
al.\ 2006, \apj, 648, 7 

\bibitem[Kennicutt(1998)]{K98} Kennicutt, R.~C., Jr.\ 1998, 
\araa, 36, 189 

\bibitem[Kodaira et al.(2003)]{Ko03} Kodaira, K., et al.\ 
2003, \pasj, 55, L17 

\bibitem[Kunth et al.(1998)]{Kunth98} Kunth, D., Mas-Hesse, 
J.~M., Terlevich, E., Terlevich, R., Lequeux, J., \& Fall, S.~M.\ 1998, 
\aap, 334, 11 

\bibitem[Lai et al.(2007)]{Lai07} Lai, K., Huang, J.-S., 
Fazio, G., Cowie, L.~L., Hu, E.~M., \& Kakazu, Y.\ 2007, \apj, 655, 704 

\bibitem[Loeb et al.(2005)]{L05} Loeb, A., Barkana, R., \& 
Hernquist, L.\ 2005, \apj, 620, 553 

\bibitem[Malhotra \& Rhoads(2002)]{MR02} Malhotra, S., \& 
Rhoads, J.~E.\ 2002, \apjl, 565, L71 

\bibitem[Malhotra et al.(2003)]{Malhotra03} Malhotra, S., Wang, 
J.~X., Rhoads, J.~E., Heckman, T.~M., \& Norman, C.~A.\ 2003, \apjl, 585, 
L25 

\bibitem[Mao et al.(2007)]{Mao07} Mao, J., Lapi, A., Granato, G.~L., de Zotti, G., \& Danese, L.\ 2007, Submitted to ApJ, astro-ph/0611799 

\bibitem[Nagao et al.(2007)]{Nagao07} Nagao, T., et al.\ 2007, 
ArXiv Astrophysics e-prints, arXiv:astro-ph/0702377 

\bibitem[Neufeld(1991)]{Neufeld91} Neufeld, D.~A.\ 1991, \apjl, 
370, L85 

\bibitem[Partridge \& Peebles(1967)]{PP67} Partridge, R.~B., 
\& Peebles, P.~J.~E.\ 1967, \apj, 147, 868 

\bibitem[Pirzkal et al.(2007)]{Pirzkal07} Pirzkal, N., Malhotra, 
S., Rhoads, J.~E., \& Xu, C.\ 2006, ArXiv Astrophysics e-prints, 
arXiv:astro-ph/0612513 

\bibitem[Press \& Schechter(1974)]{PS} Press, W.~H., \& 
Schechter, P.\ 1974, \apj, 187, 425 

\bibitem[Rhoads et al.(2000)]{Rhoads00} Rhoads, J.~E., Malhotra, 
S., Dey, A., Stern, D., Spinrad, H., \& Jannuzi, B.~T.\ 2000, \apjl, 545, 
L85 

\bibitem[Scannapieco et al.(2003)]{Sc03} Scannapieco, E., 
Schneider, R., \& Ferrara, A.\ 2003, \apj, 589, 35 

\bibitem[Schaerer(2003)]{S03} Schaerer, D.\ 2003, \aap, 
397, 527 

\bibitem[Schaerer \& Pell{\'o}(2005)]{SP05} Schaerer, D., \& 
Pell{\'o}, R.\ 2005, \mnras, 362, 1054 

\bibitem[Schneider et al.(2006)]{S06} Schneider, R., 
Salvaterra, R., Ferrara, A., \& Ciardi, B.\ 2006, \mnras, 369, 825 

\bibitem[Shapley et al.(2003)]{Shapley03} Shapley, A.~E., 
Steidel, C.~C., Pettini, M., \& Adelberger, K.~L.\ 2003, \apj, 588, 65 

\bibitem[Sheth et al.(2001)]{ST} Sheth, R.~K., Mo, H.~J., 
\& Tormen, G.\ 2001, \mnras, 323, 1 

\bibitem[Shimasaku et al.(2006)]{Shima06} Shimasaku, K., et 
al.\ 2006, \pasj, 58, 313 

\bibitem[Spergel et al.(2007)]{Spergel06} Spergel, D.~N., et al.\ 
2007, \apjs, 170, 377 

\bibitem[Stanway et al.(2004)]{Stanway04} Stanway, E.~R., et al.\ 
2004, \apjl, 604, L13 

\bibitem[Stanway et al.(2005)]{Stanway05} Stanway, E.~R., 
McMahon, R.~G., \& Bunker, A.~J.\ 2005, \mnras, 359, 1184 

\bibitem[Stanway et al.(2007)]{Stanway07} Stanway, E.~R., et al.\ 
2007, \mnras, 376, 727 

\bibitem[Stark et al.(2007)]{Stark07} Stark, D.~P., Loeb, A., 
\& Ellis, R.~S.\ 2007, ArXiv Astrophysics e-prints, arXiv:astro-ph/0701882 

\bibitem[Taniguchi et al.(2005)]{Taniguchi05} Taniguchi, Y., et 
al.\ 2005, \pasj, 57, 165 

\bibitem[Tapken et al.(2007)]{Tapken07} Tapken, C., Appenzeller, 
I., Noll, S., Richling, S., Heidt, J., Meink{\"o}hn, E., \& Mehlert, D.\ 
2007, \aap, 467, 63 

\bibitem[Tumlinson et al.(2001)]{T01} Tumlinson, J., 
Giroux, M.~L., \& Shull, J.~M.\ 2001, \apjl, 550, L1 

\bibitem[Tumlinson et al.(2003)]{T03} Tumlinson, J., Shull, 
J.~M., \& Venkatesan, A.\ 2003, \apj, 584, 608 

\bibitem[Verhamme et al.(2006)]{V06} Verhamme, A., 
Schaerer, D., \& Maselli, A.\ 2006, \aap, 460, 397 

\bibitem[Wang et al.(2004)]{Wang04} Wang, J.~X., et al.\ 2004, 
\apjl, 608, L21 

\bibitem[Westra et al.(2006)]{Westra06} Westra, E., et al.\ 
2006, \aap, 455, 61 

\bibitem[Wyithe \& Cen(2007)]{WC} Wyithe, J.~S.~B., \& 
Cen, R.\ 2007, \apj, 659, 890 

\bibitem[Wyithe \& Loeb(2007)]{WL07} Wyithe, J.~S.~B., \& 
Loeb, A.\ 2007, \mnras, 375, 1034 

\end{thebibliography}
\end{document}